\documentclass[conference]{IEEEtran}

\usepackage{cite}
\usepackage{amsmath,amssymb,amsfonts}
\usepackage{algorithmic}
\usepackage{textcomp}
\usepackage[table]{xcolor}
\usepackage{graphicx}
\usepackage{url}

\usepackage{dblfloatfix}
\usepackage{tabularx}
\usepackage{booktabs}
\usepackage{multirow}
\usepackage{colortbl}

\def\BibTeX{{\rm B\kern-.05em{\sc i\kern-.025em b}\kern-.08em
    T\kern-.1667em\lower.7ex\hbox{E}\kern-.125emX}}

\title{Automating Quality Assessment of LLM-Generated Defeaters}

\author{
T.~Rohlinger,
D.~Ratiu,
and S.~Wagner
}

\begin{document}

\maketitle

\begingroup
\renewcommand\thefootnote{}
\footnotetext{\copyright~2025 IEEE. Personal use of this material is permitted. 
Permission from IEEE must be obtained for all other uses, in any current or future media, 
including reprinting/republishing this material for advertising or promotional purposes, 
creating new collective works, for resale or redistribution to servers or lists, or reuse 
of any copyrighted component of this work in other works. Published version: 
doi:10.1109/ICSRS68021.2025.11422208.}
\endgroup

\begin{abstract}
High-integrity systems, such as autonomous vehicle fleets or large-scale energy infrastructures, rely on structured assurance cases, represented as rooted directed acyclic graphs (DAGs), to justify safety claims. To remain valid in the face of change, these cases must be robust against potential challenges, known as 'defeaters'. While large language models (LLMs) have recently enabled scalable generation of such defeaters, validating their quality remains a predominantly manual and subjective process. This paper presents an automated method for assessing LLM-generated defeaters using natural language processing (NLP) techniques. Our approach combines structural analysis of assurance case graphs with vector-based semantic embeddings and meta-classifiers trained on expert-assessed consensus defeaters. We evaluate our method through two case studies in the automotive and energy domains, quantifying human reviewer dissensus using Cohen’s kappa (\( \kappa < 0.442 \)), which indicates low inter-rater agreement. Our automated approach achieves greater consistency with individual raters, improving (\( \kappa \approx 40\% \)). The method delivers an average F1-score of 0.84 across validation, reducing subjective variance through scalable, objective assessment. Our approach aims to advance the tool support for automation of assurance case synthesis.

\end{abstract}
\section{Introduction}\label{sec:Introduction}
\begin{figure}[ht]
    \centering
    \includegraphics[scale=0.5]{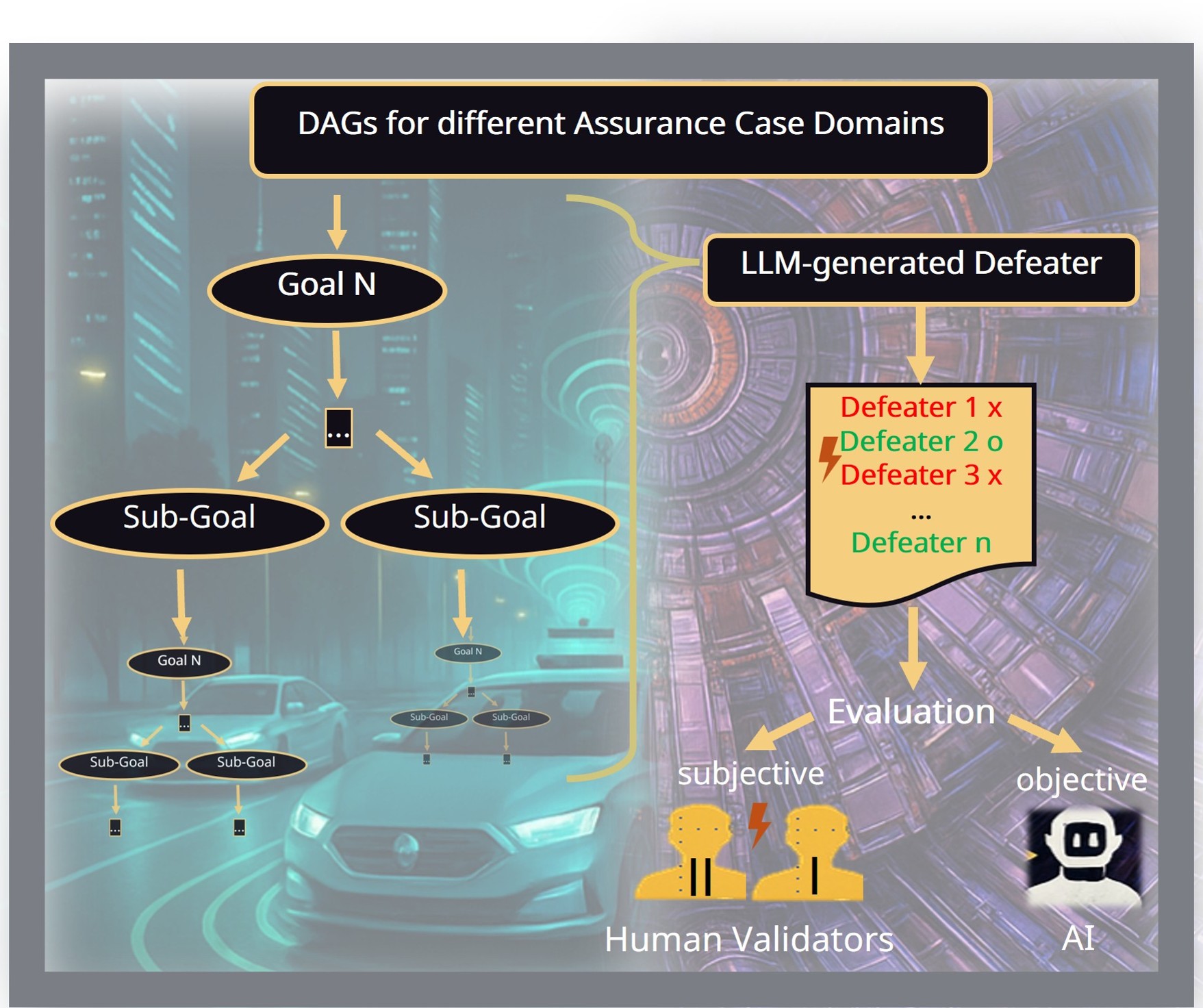}
    \caption{This is an assurance case graph in a goal-structured notation style \cite{Kelly2004}, showing goals and sub-components as nodes (left). The diagram illustrates the manual reevaluation process of the generated defeaters and proposed automated AI assessment to mitigate subjective bias (right). We evaluated the approach on two assurance case studies (background), the Adaptive Cruise Control (ACC) (left) from \cite{viger_ai-supported_2024} and Large Hadron Collider (CERN LHC) (right) from \cite{millet_assurance_2023}.}
    \label{fig:Overview_total}
\end{figure}

The research community aims to establish a validation method for safety case fragments, assuring over time-changing safety-critical systems that adapt to evolving safety boundaries \cite{ratiu_towards_2024,koopman_how_2022}. The pace of adaptation required is determined by the exposure of entities in, for example, autonomous driving fleets. These kinds of systems demand robust safety cases, enhanced by runtime monitoring feedback \cite{2015_Assadi}, thereby enabling advanced system-level assurance \cite{tihomir_automated_2024}.
Defeaters, in the domain of safety assurance, refer to arguments or evidence that undermine the assurance claims made within the safety case \cite{Bloomfield-etal:defeaters24}. Identifying and understanding these defeaters are essential for validating a safety case's resilience against real-world scenarios. Manual creation and validation in this context is not scalable and relies on subjective expert judgment \cite{rushby_interpretation_2015}. Generating defeaters using large language models (LLMs) increases confidence in the argumentation \cite{viger_ai-supported_2024}, thereby making safety assurance scalable and responsive.\\ In Viger et al. (2024) \cite{viger_ai-supported_2024}, the generated defeaters were manually reviewed by two safety experts. The experts only agreed on the quality of the defeaters, with a low level of inter-rater agreement calculated by cohens kappa \cite{everitt_encyclopedia_2005}.
Low inter-rater agreement indicates that expert judgments vary due to differing interpretations and domain knowledge. Furthermore, manual validation is resource-intensive and lacks scalability, particularly for dynamic systems with evolving safety boundaries, such as adaptive software. These limitations hinder the efficiency and reliability of assurance synthesis, making automated approaches necessary to enhance defeater validation. This work addresses the issue of subjectivity in manual defeater validation, as illustrated by the schematic figure \ref{fig:Overview_total}. \\
In Section \ref{sec:Method}, we propose an NLP-based method that uses BERT embeddings \cite{devlin_bert_2019} and meta-classifiers to objectively evaluate LLM-generated defeaters, enabling scalable assurance. Our experiment, detailed in Section \ref{sec:Exp}, compares this approach with human evaluations and demonstrates improved consistency and reduced subjectivity in expert dissent. The results in Section \ref{sec:Results} provide further support for the conclusions in Section \ref{sec:Con} with an outlook on future work.

\section{Related Work}\label{sec:Lit}
\textbf{LLM-generated Assurance Fragments}
Recent studies, such as Sivakumar et al. (2023) \cite{sivakumar_gpt-4_2023}, demonstrated GPT-4’s moderate success in generating Goal Structuring Notation (GSN)-based safety cases, a graph structure style, limited by incomplete domain knowledge \cite{sivakumar_prompting_2024}. Kurzidem et al. (2023) \cite{kurzidem_safesens_2023} used meta-models to enable a holistic safety analysis to quantify the uncertainties. Gohar et al. (2024) \cite{gohar_codefeater_2024} showed preliminary success in using LLMs to identify defeaters, while Shahandashti et al. (2024) \cite{khakzad_shahandashti_assessing_2024} highlighted GPT-4 Turbo’s capability to generate domain-specific defeaters like software bugs or environmental risks, thereby further advancing automation.
Graydon (2025), appendix, finds the redefinition of defeaters as hazards or failure conditions, distinct from logical flaws, suggesting that generic defeater searches may not replace safety analysis \cite{graydon_examining_2025}.
Defeaters are reasons that challenge a claim through defeasible reasoning \cite{goodenough_eliminative_2015}.

\textbf{Embeddings for Assurance Cases Fragments}
The Transformer architecture \cite{vaswani_attention_2017} has revolutionized natural language processing by enabling large language models (LLMs), such as GPT-4, to generate contextual embeddings that capture intricate semantic and syntactic relationships \cite{devlin_bert_2019}. These embeddings serve as a foundation for advanced techniques like Named Entity Recognition (NER) \cite{Keraghel2024RecentAI}. By leveraging the contextual richness of LLM-generated embeddings, assurance case development can produce clear, domain-specific narratives within frameworks like Goal Structuring Notation (GSN), ensuring both structural integrity and semantic accuracy \cite{sivakumar_prompting_2024}. Existing research often overlooks automated quality assessments of generated claims and the use of advanced classification techniques to validate them. Recent studies that apply LLMs to the analysis of safety-critical systems include goal modeling \cite{chen_use_2023}, automated software proof generation \cite{first_baldur_2023}, and LLM-driven defeater identification \cite{viger_ai-supported_2024}. Odu et al. (2025) \cite{Odu2025AutomaticIO} compared LLM outputs and assurance cases, finding that outputs are "relatively good" but not sufficient for certifying mission-critical systems without human intervention. Varadarajan et al. (2024) \cite{varadarajan_enabling_2024} propose a continuous assurance framework that emphasizes validity and soundness, employing LLMs to extract properties and environmental information from assurance case texts, processed by Prolog-based reasoning engines for semantic and logical analysis. While effective for structured validation, such hybrid approaches often rely on LLMs for natural language syntactic parsing rather than direct context semantic analysis. 

\textbf{Verification and Validation}
Eliminative Argumentation (EA) in this regard provides a structured approach to constructing and validating assurance arguments by eliminating defeaters \cite{goodenough_eliminative_2015}. This iterative process builds confidence in claims by systematically addressing doubts, making it suitable for automation through NLP.
Ensuring LLM reliability requires robust validation against real-world data, expert evaluations, and alignment with standards like ISO 26262, ISO 21448, and UL4600 \cite{iso_26262_1_2018, ISO21448, ansi_ul_4600_2023}. These standards enhance the administration of domain-specific safety case generation and independent body compliance, confirming stakeholder acceptance. Rushby et al. \cite{rushby_interpretation_2015} emphasize evaluating assurance arguments beyond compliance, advocating for semantic robustness. We assume that semantic robustness can be enhanced through precise claim localization in DAGs and objective defeater embedding analysis by NLP techniques. Validation strategies, as discussed in \cite{turobov_using_2024}, address risks such as output inaccuracy, lack of systematic coding frameworks, and ethical concerns in the validation and verification of LLM-generated code, emphasizing the need for rigorous processes to ensure reliable and safe outcomes. An iterative process builds confidence \cite{rushby_interpretation_2015} in safety-critical systems, such as autonomous driving, by addressing doubts to support claims like “the system is safe”.
Viger et al. (2024) \cite{viger_ai-supported_2024} leveraged EA to generate defeaters for assurance cases, providing input for our work. Building on the defeaters generated by Viger et al., our approach uses the BERT architecture \cite{devlin_bert_2019} to automate their evaluation and validates it with a trained model on expert judgment. This case study lays the foundation for experiments that can automatically evaluate LLM-generated defeater performance in safety-critical contexts.

\textbf{Summary}
Existing studies focus on generating assurance case fragments but do not address the systematic evaluation of their quality, leaving a gap in reliable and objective validation. In this work, we focus on automating the quality assessment of synthesized assurance arguments and the use of classifier strategies to validate defeaters.

\section{Method}\label{sec:Method}
This section presents the methodology for assessing the quality of defeaters within assurance cases using Natural Language Processing (NLP) techniques. 
Our methodology extends the AI-Supported Eliminative Argumentation (AI-EA) framework by Viger et al.\cite{viger_ai-supported_2024} comprising three phases: (1) the foundation experiment establishing a baseline through manual review, we will present a small insights from the provided dataset from Viger et al. to clarify the manual review task; (2) an inter-rater agreement analysis to quantify consensus and dissensus of the manual review; and (3) our novel extention, the automated review process.

\subsection{Manual Review Process}\label{subsec:Manual Review Process}

\begin{table*}[!b]
    \centering
    \smallskip
    \smallskip
    \caption{Defeater Evaluation Metrics for Quality Assessment using NLP}
    \begin{tabular}{|p{1.9cm}|p{7.2cm}|p{7.2cm}|}
        \hline
        \textbf{Attributes} & \textbf{Interpretation Description} & \textbf{Metric Evaluation based on NLP} \\
        \hline
        \textbf{Correctness (C)} & Evaluates how accurately the expert reviews match the other judgments in rating components ("What", "Where", and "Why"). & F1 scores assess the accuracy of each class (0, 1 and 2) in comparison to human reviewers H1 and H2. They measure the ability of the models to correctly classify the defeater labels. \\
        \hline
        \textbf{Stability (St)} & Associates consistent labels or probabilities for defeaters with high semantic similarity, ensuring stable recognition for defeaters with closely related meanings. & The proportion of identical predicted labels for defeater pairs with cosine similarity using BERT embeddings. Consistent labels across similar defeaters indicate robust performance. \\
        \hline
        \textbf{Relevance (R)} & Ensures that defeaters are applied to the assurance case and that relevant claims contribute meaningfully to the system’s reliability evaluation. & The average cosine similarity between the BERT embeddings of the defeaters and the elements of the assurance case graph (e.g. goal claims, strategies etc. DAG edges) indicates the degree of alignment with the case.\\
        \hline
        \textbf{Completeness (Comp)} & Reviewer checks whether defeaters include necessary components (What, Where, Why) to address issues within the assurance case, ensuring comprehensive and contextually complete defeaters. & The proportion of defeaters with non-empty What and Why components is extracted via text parsing. The Where component is covered by a regular expression filter, ensuring logical and contextual coverage within the assurance case hierarchy. \\
        \hline
        \textbf{Novelty (N)} & Evaluates whether defeaters introduce new insights or challenges to avoid redundancy with existing defeaters and to promote originality in identifying potential issues. & Inverse of the minimum cosine similarity between a defeater’s BERT embedding and those of reference defeaters, where lower similarity suggests greater novelty and reduced redundancy. \\
        \hline
    \end{tabular}
    \label{tab:comp_metric}
\end{table*}

The AI-Supported Eliminative Argumentation (AI-EA) study dataset includes 171 AI-generated defeaters from two industrial assurance cases (ACs): an Adaptive Cruise Control (ACC) system from the automotive domain including (317 nodes) and the CERN Large Hadron Collider Machine Protection System (MPS) from the nuclear domain with (506 nodes) \cite{Dataset}. 
Each generated defeater was structurally analyzed as a textual counterargument within the Eliminative Argumentation framework \cite{goodenough_eliminative_2015}, comprising three components: ``What'' (the flaw), ``Where'' (the affected claim or system component), and ``Why'' (the rationale for the challenge). Reviewers scored these components as 0 (absent or non-compliant), 1 (present but generic), or 2 (specific and reasonable). Addressing defeaters strengthens safety arguments, thereby enhancing AC confidence. Below, we provide two example defeaters to illustrate their structure and the manual analysis process, as detailed in the study by Viger et al..
\begin{itemize}
    \item \textbf{ACC Defeater ID 1}: ``\textit{The effectiveness of the black box radar sensing and Machine Learning system in maintaining safe distance is uncertain; Where: reliance on a system whose internal functioning is not analyzed; Why: without understanding the system's internal functioning, potential failure modes may not be predictable or preventable.}''\\
    \textit{Analysis}: Rated 1 (generic) for ``What'' (vague uncertainty), 0 (non-compliant) for ``Where'' (lacks specific claim reference), and 2 (reasonable) for ``Why'' (clear failure mode rationale). This defeater prompts analysis of system behavior to address black box uncertainties.
    
    \item \textbf{CERN Defeater ID 1}: ``\textit{What: Total reliance on the Beam Loss Monitoring System (BLMS) for detecting intolerable beam loss; Where: CHILD CLAIM 9; Why: assumes BLMS always provides timely indications, risking system failure if BLMS faults.}''\\
    \textit{Analysis}: Rated 2 (reasonable) for ``What'' and ``Where'' (specific system and claim), but 1 (generic) for ``Why'' (broad fault assumption). This defeater flags a single point of failure, necessitating reliability evidence.
\end{itemize}
These examples demonstrate how defeaters were manually analyzed individually for component quality and how addressing them (e.g., through evidence or refined claims) strengthens safety case confidence.

\subsection{Consensus and Dissensus}\label{subsec:Consensus and Dissensus}
Following the manual review process described in Section \ref{subsec:Manual Review Process}, we assessed the inter-rater agreement between the two human reviewers (H1 and H2) to quantify the consistency of defeater ratings for the ``What,'' ``Where,'' and ``Why'' components across the ACC and CERN datasets. This analysis establishes the baseline variability in expert judgments, motivating the need for automated validation. We calculated Cohen’s kappa categorical using the \textit{cohen\_kappa\_score} function from the \textit{scikit-learn} library \cite{scikit-learn}, accounting for agreement by chance. The methodology involved the following steps:

\begin{enumerate}
    \item \textbf{Observed Agreement (\(p_o\))}: Computed as the proportion of defeaters where H1 and H2 assigned the same rating (0, 1, or 2) for each component.
    \item \textbf{Marginal Probabilities}: Probabilities that each rater assigns a specific rating to a component, regardless of the other rater’s choice. Obtained from the rating distributions.
    \item \textbf{Expected Agreement (\(p_e\))}: Calculated as the sum of the products of corresponding marginal probabilities for each rating level.
    \item \textbf{Cohen’s Kappa (\(\kappa\))}: Determined using the formula:
    \[
    \kappa = \frac{p_o - p_e}{1 - p_e}
    \]
    as described in \cite{everitt_encyclopedia_2005}.
\end{enumerate}
For example, the ACC dataset with the ``Why'' component, reviewers agreed on 42 out of 75 defeaters, with an observed agreement (\(p_o = 0.56\)), and an expected agreement of (\(p_e \approx 0.57\)). The rating distributions were:
\begin{itemize}
    \item Reviewer 1: \{0: 1, 1: 19, 2: 55\}
    \item Reviewer 2: \{0: 1, 1: 24, 2: 50\}
\end{itemize}
The kappa values were:
\begin{itemize}
    \item \textbf{What}: \(\kappa = 0.383\), indicating fair agreement.
    \item \textbf{Why}: \(\kappa = -0.024\), implying disagreement below chance.
    \item \textbf{Where}: \(\kappa = 0.442\), indicating moderate agreement.
\end{itemize}
For the CERN dataset, the ``Why'' component, here the reviewers agreed on 61 out of 97 defeaters (\(p_o \approx 0.629\)), with \(p_e \approx 0.565\). The rating distributions were:
\begin{itemize}
    \item Reviewer 1: \{1: 32, 2: 65\}
    \item Reviewer 2: \{0: 2, 1: 26, 2: 69\}
\end{itemize}
The kappa values were:
\begin{itemize}
    \item \textbf{What}: \(\kappa = 0.167\), indicating slight agreement.
    \item \textbf{Why}: \(\kappa = 0.147\), indicating slight agreement.
    \item \textbf{Where}: \(\kappa = 0.329\), indicating fair agreement.
\end{itemize}

These results highlight significant inter-rater variability, particularly the negative kappa for the ACC ``Why'' component, underscoring the subjectivity in manual evaluations. This variability underscores the need for an automated, structured review process that follows qualitative, objective rating metrics. This will help reduce bias and improve consistency.

\begin{figure*}[b]
    \centering
    \includegraphics[scale=0.6]{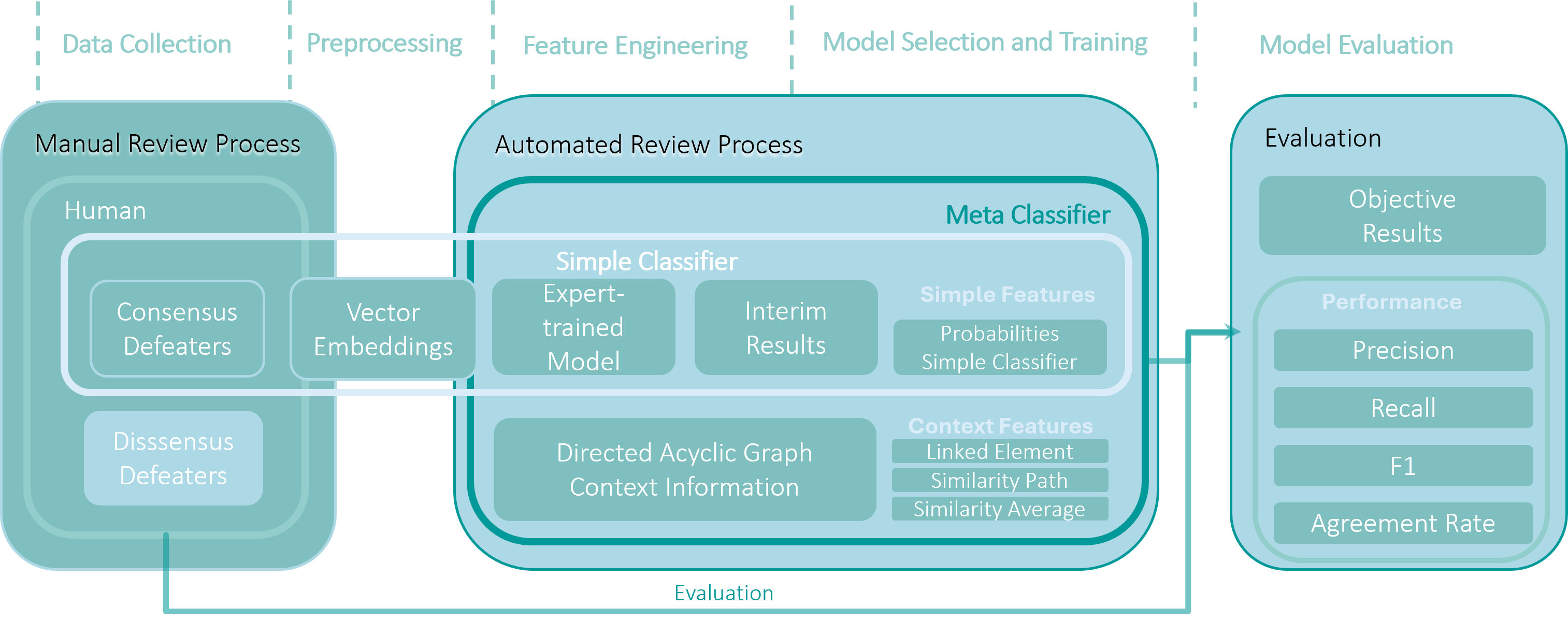}
    \caption{Automated review process and classifier integration}
    \label{fig:Automated_Review_Process}
\end{figure*}

\subsection{Quality Assessment of Defeaters} \label{subsec:Quality Assessment of Defeaters}
As discussed in Subsection \ref{subsec:Consensus and Dissensus}, manual validation of defeaters often exhibits significant inter-rater variability, reflecting inconsistencies in the subjective judgements of evaluators. This variability undermines the reliability and reproducibility of assurance case assessments, particularly in safety-critical domains where objective rigour is paramount. To address these issues, we have formalised a set of quality attributes for defeaters derived from established manual evaluation practices. These attributes are designed to be machine-processable, enabling standardised, automated assessment. 
Table \ref{tab:comp_metric} presents these attributes (left), their Interpretation Description (center), and how these attributes can be quantified by our NLP Metric (right) to enable machine-processable assessment. The attributes are: \textit{Correctness} aligns automated ratings with human judgments using F1-scores. \textit{Stability} ensures consistent labels for similar defeaters via BERT embeddings' cosine similarity. \textit{Relevance} reflects the defeater's affiliation within the assurance case graph. \textit{Completeness} verifies component presence through text parsing, informing field detection. \textit{Novelty} assesses uniqueness with inverse similarity, diversifying features. This variability-driven approach shapes our automated review design, enhancing the reflective NLP features extraction for the machine learning models.

\subsection{Automated Review Process} \label{subsec:Automated Review Process}
From the identification of the machine-processable quality attributes of the previous subsection \ref{subsec:Quality Assessment of Defeaters}, we derived the Automated Review Process.
Figure \ref{fig:Automated_Review_Process} illustrates this process chart with three primary nodes: Manual Review Process, Automated Review Process, and Evaluation, arranged to depict the transition from human assessment to automated validation and its performance assessment. The details of each node and its subcomponents are provided below, along with their positions and functionalities within a machine-learning cycle. This cycle is clustered on the upper rim of the process chart figure.

\subsubsection{Data Collection}
The Manual Review Process, positioned as the leftmost node in Figure \ref{fig:Automated_Review_Process}, encompasses the initial human-reviewed defeater data mentioned in subsection \ref{subsec:Manual Review Process}, comprising the defeaters for each component. The Human node includes two sub-categories: \textit{Consensus Defeaters}, where reviewers agree (H1=H2), and \textit{Dissensus Defeaters}, where they disagree (H1xH2), reflecting inter-rater variability as discussed in section Consensus and Dissensus \ref{subsec:Consensus and Dissensus}. This bridges to the preprocessing node of the Automated Review Process, the central node in Figure \ref{fig:Automated_Review_Process}.

\subsubsection{Preprocessing}
The embedding preprocessing employs the \textit{bert-base-uncased} model \cite{DBLP:journals/corr/abs-1810-04805} from Hugging Face, a transformer-based architecture with 12 layers and a 768-dimensional output representation, as the initial choice. To generate embeddings for defeaters, the input text is tokenized with a maximum sequence length of 512 tokens. The token-level embeddings produced by the model are aggregated into a single vector representation for each defeater by computing the mean across the token dimension.
For creating the semantic relation of embeddings and the top-down hierachical structure, a directed graph of assurance case elements was constructed. This included nodes representing claims, evidence, strategies, contexts, or defeaters, and edges denoting hierarchical dependencies, facilitating the contextual analysis. This data preparation underpins the feature engineering.

\subsubsection{Feature Engineering}\label{subsec:Feature Engineering}
Validity features were engineered to enhance classification, derived from the quality attributes outlined in section \ref{subsec:Quality Assessment of Defeaters} Table \ref{tab:comp_metric}. These features support the automated validation process, interpreting the expert judgment in a machine-quantifiable format. Thereby, it provides a transparent, reproducible factor to enable objective evaluation. 
\begin{itemize}
    \item \textbf{Is Linked}: A binary indicator (1 if the defeater references a linked assurance case element, 0 otherwise), derived from the Relevance and Completeness attribute to assess applicability to the case claims.
    \item \textbf{Linked Element}: Cosine similarity between the defeater’s BERT embeddings and the embedding of its linked assurance case element (if present, Is Linked=1), supporting Relevance by quantifying semantic alignment.
    \item \textbf{Similarity Path}: Cosine similarity between the defeater’s BERT embedding and the concatenated text of the shortest path from the assurance case graph’s root to the linked element. Thereby reflecting the Stability attribute by ensuring consistent predictions for related elements.
    \item \textbf{Similarity Average}: Mean cosine similarity between the defeater’s BERT embedding and the assurance case embeddings, enhancing Relevance by capturing overall contextual fit and its opposite Novelty to define whether defeaters are novel.
\end{itemize}
Following, we will introduce the models that were trained on the features presented. 

\subsubsection{Model Selection and Training}\label{subsec:Model Selection and Training}
Two classification approaches were implemented using \textit{scikit-learn}:\begin{itemize}[\vspace{\baselineskip}]
\item \textbf{Simple Classifier}: The Logistic Regression (LR) employs multinomial softmax regression with L2 regularization for multiclass classification and balanced class weights compensating for data imbalances. LR serves as a baseline or complementary model. Secondary Support Vector Machines (SVMs) that find the optimal hyperplane to separate classes in a feature space were trained on BERT embeddings for the What and Why components. The hyperparameters (kernel: \{linear, rbf\}, C: \{0.1, 1, 10, 100\}, gamma: \{scale, 0.001, 0.01, 0.1\}) were tuned using \textit{GridSearchCV} (e.g., all combinations of kernel, C, gamma) with 5-fold stratified cross-validation (or fewer if class sizes were small ex. Why 0 class ratings), optimizing for weighted F1-score. These hyperparameters are tuned to find the best SVM configuration for classifying defeater components, balancing fit and generalization. Class imbalance was addressed using class weights, and probabilities were calibrated with\\ \textit{CalibratedClassifierCV} (sigmoid method), to better reflect true confidence levels.
\item \textbf{Meta-Classifier}: LR and SVMs combinations were trained on a feature set combining simple classifier probabilities and validity features (is linked, linked element, path similarity, similarity average) as introduced in subsection Feature Engineering \ref{subsec:Feature Engineering}. The same hyperparameter tuning and calibration process was applied as for the Simple Classifier. To address class imbalance, the Synthetic Minority Oversampling Technique \textit{SMOTE} \cite{Chawla2002SMOTE} was used. Additionally, a self-training approach using the SelfTrainingClassifier \cite{yarowsky-1995-unsupervised} with a confidence threshold of 0.9 was applied. This threshold determines the minimum confidence required to assign pseudo-labels to unlabeled data, reducing the risk of incorporating incorrect labels, enhancing model robustness.
\end{itemize}\vspace{\baselineskip}

\subsubsection{Evaluation}
With the method section \ref{sec:Method} establishing the automated review process, we will apply it to the ACC and CERN datasets \cite{Dataset}, evaluating the performance of the models based on the \textit{scikit-learn}, providing F1 scores and \textit{cohen\_kappa\_score} Agreement Rates. 

Improving objectivity in the task performance by leveraging the graph-based embeddings for consistency, to prove the feasibility of AI-aided defeater validation for assurance synthesis.

\section{Experiment}\label{sec:Exp}
This experiment addresses the context gap between manual and automated validation of generated textual output, specifically defeaters, which are core elements in the argumentation of system safety. Defeaters require subjective review due to their role in challenging safety claims, and this study seeks to recreate such validation using Natural Language Processing (NLP) methods as elaborated in \ref{tab:comp_metric}, to achieve objective, bias-independent results. The forthcoming experiments extend the foundational research executed by Viger et al. \cite{viger_ai-supported_2024}, which supplies the ACC and CERN assurance cases as well as their corresponding defeater datasets (ACC and CERN) \cite{Dataset}.

\subsection{Hypotheses and Research Questions}\label{subsec:Hypotheses}
We hypothesize that a vector-based embeddings approach outperforms subjective expert reviews as outlined in \cite{viger_ai-supported_2024}, in terms of consistency (measured by F1-score) and objectivity (measured by agreement on dissensus defeaters). This addresses the inter-rater variability observed in Section \ref{subsec:Consensus and Dissensus}. The experiment investigates two research questions:
\begin{enumerate}
    \item \textbf{RQ1}: Is the validation of defeaters using an embedding vector-based method comparable to human expert evaluations in terms of accuracy and consistency?
    \item \textbf{RQ2}: Does the embedding-based approach reduce subjectivity by providing an objective perspective on safety case defeater ratings?
\end{enumerate}

\subsection{Experimental Setup}\label{subsec:Setup}
The data \cite{Dataset} was obtained from the experimental results of Viger et al. \cite{viger_ai-supported_2024}, in which two reviewers collaborated with domain experts to independently analyse the 172 defeaters (75 for ACC and 97 for CERN) generated by GPT-4. These defeaters, covering nine argument fragments, were prompted by three strategies: background information, generic doubts, and context-specific challenges informed by Eliminative Argumentation (EA) \cite{goodenough_eliminative_2015}. Each defeater comprises ``What'' (the flaw), ``Why'' (the rationale), and ``Where'' (the affected claim or system component). The automated validation rates the ``What'' and ``Why'' components as 0 (non-compliant), 1 (generic), or 2 (reasonable), using expert annotations as the ground truth.

The methodology, detailed in Section \ref{subsec:Automated Review Process}, employs open-source BERT embeddings (``bert-base-uncased'') \cite{devlin_bert_2019} to transform defeater texts and assurance case structures into 768-dimensional vectors. Logistic Regression (LR), Support Vector Machine (SVM), and meta-SVM classifiers predict ratings. Therby incorporating defeater text and structural context in the DAGs (e.g., claim hierarchies via \textit{networkx}). The experiment proceeds in two rounds: (1) using defeater text alone, and (2) integrating structural context via features to enhance predictions. Performance is evaluated against dissensus annotations (H1$\neq$H2) using precision, recall, and F1-score, addressing the quality attributes in Table \ref{tab:comp_metric}.

\subsection{Data Preprocessing}\label{subsec:DataPreprocessing}
The ACC and CERN datasets, stored in \textit{ACC-CERN\_defeater\_dataset.xlsx} \cite{Dataset}, contain 172 defeaters annotated by two human raters (H1, H2) for ``What'', ``Why'', and ``Where'' components on a \{0, 1, 2\} scale, yielding 688 samples (172 defeaters $\times$ 2 reviewers $\times$ 2 components). The dataset includes columns for Defeater ID, Defeater text, component ratings, and rater information. We split the data into:\\
\begin{itemize}
    \item \textbf{Consensus Sets}: Defeaters where H1 and H2 agree on ``What'' and ``Why'' ratings (H1=H2). For ACC, 32 defeaters (31 initial, plus 1 dissensus for class balance); for CERN, 42 defeaters (40 initial, plus 2 dissensus).
    \item \textbf{Dissensus Sets}: Defeaters with mismatched ratings (H1$\neq$H2) or insufficient samples, used for evaluation (42 for ACC, 50 for CERN after transfers to consensus sets).
\end{itemize}
To address class imbalance in the training sets, we applied SMOTE \cite{Chawla2002SMOTE} for the ``What'' and ``Why'' components. For each dataset, if a class had fewer than 5 samples, SMOTE was applied with a \( k \)-nearest neighbors parameter set to \(\min(3, n-1)\) a common default as it balances local information and generalization. Where \( n \) is the number of samples in the minority class. If $ n = 1 $, SMOTE cannot be applied (since $ k = n-1 = 0 $), and the script falls back to generating synthetic defeaters using \textit{nlpaug} with WordNet-based synonym augmentation.
The final training sets, after SMOTE and synthetic augmentation, included: for ACC, 32 consensus defeaters supplemented with up to 5 synthetic samples per underrepresented class in ``What'' and ``Why''; for CERN, 42 consensus defeaters similarly augmented. The evaluation sets remained unchanged in size, with 42 defeaters for ACC and 50 for CERN. 

For the context-informed evaluation step 2), a directed graph of assurance case elements (claims, evidence, strategies) was constructed using \textit{networkx} to capture dependencies, with nodes and edges derived from the safety case CSV files (\textit{ACC.csv}, \textit{CERN\_LHC.csv}) \cite{Dataset}. Outputs, including raw datasets, consensus and dissensus sets, and training/evaluation splits, are saved as CSV files for reproducibility.

\subsection{Embeddings and Features}\label{subsec:Features}
We use the \textit{bert-base-uncased} model \cite{devlin_bert_2019} to generate \\ 768-dimensional embeddings for defeater texts, computed as the mean of the last hidden state. Empty or invalid inputs return zero vectors. For the context-informed approach, a meta-classifier integrates the validity features introduced in section \ref{subsec:Feature Engineering}. 

\begin{itemize}
    \item \textbf{Is Linked}: Binary indicator
    \item \textbf{Cosine Similarity}: Between defeater \& linked element.
    \item \textbf{Path Similarity}: Between defeater \& path route.
    \item \textbf{Average Cosine Similarity}: Mean graph similarity.
\end{itemize}

These features are normalised using min-max scaling and combined with the probabilities of the simple classifiers reassembled for the meta-classifier predictions, as illustrated in the 'meta-classifier node' of Figure \ref{fig:Automated_Review_Process} in the method section \ref{sec:Method}.

\subsection{Model Description}\label{subsec:ModelDescription}
The experiment trains the two types of classifiers detailed in Subsection \ref{subsec:Model Selection and Training} for ``What'' and ``Why'' components:
\begin{itemize}
    \item \textbf{Simple Classifiers}: LR and SVM models predict labels \{0, 1, 2\} using BERT embeddings of defeater texts.
    \item \textbf{Meta-Classifiers}: LR and SVM meta-classifiers refine predictions by integrating simple classifier probabilities and validity features as mentioned in the paragraph above. Labels are encoded with \textit{LabelEncoder}, and min 2-fold cross-validation ensures balanced data training.
\end{itemize}
Feature importance is computed via coefficients for linear kernels and permutation importance for non-linear kernels.

\subsubsection{Extraction Logic}\label{Subsec:Extraction Logic}
Defeaters were matched to safety case nodes using regex to identify ID-based references (e.g., "Child Claims 5000", "Context 3001"). The logic targeted terms like "Text Strategy", "Text Context", "Parent", "Child", "Evidence", or numeric IDs in the "Where" field. Defeaters with descriptive 'Where' fields lacking structured references were rated 0 for positioning due to insufficient linkage to safety case nodes and were excluded from semantic analysis. Positional verification was performed by searching against the related safety case claims.

\subsection{Evaluation}\label{subsec:Evaluation}
The evaluation is conducted in two rounds:
\begin{enumerate}
    \item \textbf{Text-Only}: Simple classifiers predict ratings using only defeater text embeddings.
    \item \textbf{Context-Informed}: Meta-classifiers incorporate structural DAG context (validity features) to improve predictions.
\end{enumerate}

Performance is assessed against dissensus annotations (H1$\neq$H2) using precision, recall, and weighted F1-score, as defined in Table \ref{tab:comp_metric}. The weighted F1 score is a metric that computes the F1 score for each class and then averages them, weighted by the number of samples in each class (support). Correctness is evaluated by comparing predictions to H1 and H2 ratings, while Stability, Relevance, Completeness, and Novelty are computed as described in Section \ref{subsec:Quality Assessment of Defeaters}. An ablation study was conducted to evaluate the different configurations for class imbalance mitigation: baseline (SVM simple SVM Meta), SMOTE, non-consensus data inclusion, self-training, and SMOTE with non-consensus data. We followed the default random state 42 for data splitting and initialization to produce the same result base across runs.

\subsection{Limitations}\label{subsec:Limitations}
The experiment faces several limitations:
\begin{itemize}
    \item \textbf{Data Quality}: Inconsistent or incomplete annotations, particularly for the ``Where'' component, impact embedding reliability. Class imbalance in training (e.g., few ``0'' ratings) is mitigated with dissensus integration; the small sample sizes limit robustness.
    \item \textbf{Human Bias}: Variability in expertise for a domain influences annotations (Section \ref{subsec:Consensus and Dissensus}), introducing bias in the training data, which is partially addressed by focusing on consensus-defeater training and extracting objective context from the structure within the assurance case.
    \item \textbf{Model Complexity}: BERT’s 768-dimensional embeddings capture rich semantics but could be enriched with more dimensionality. The LR and SVM classifiers may oversimplify nuanced defeater semantics based on the chosen parameters.
\end{itemize}

\section{Results}\label{sec:Results}
\begin{table*}[!htbp]
\centering
\caption{Model Performance Comparison General}
\label{tab:model_comparison}
\small
\setlength{\tabcolsep}{4pt}
\begin{tabularx}{\textwidth}{p{3.5cm} *{6}{>{\centering\arraybackslash}X}}
\toprule
\textbf{Classifier} & \multicolumn{3}{c}{\textbf{What}} & \multicolumn{3}{c}{\textbf{Why}} \\
\cmidrule(lr){2-4} \cmidrule(lr){5-7}
& \makebox[2em]{Precision} & \makebox[2em]{Recall} & \makebox[2em]{F1} & \makebox[2em]{Precision} & \makebox[2em]{Recall} & \makebox[2em]{F1} \\
\addlinespace
\midrule
\multicolumn{7}{c}{\textbf{Data ACC}} \\
\midrule
Simple LR & 0.64 & 0.80 & 0.71 & 0.85 & 0.92 & 0.88 \\
Simple SVM & 0.8 & 0.82 & 0.81 & 0.84 & 0.84 & 0.84 \\
LR meta, LR simple & 0.82 & 0.80 & 0.73 & 0.85 & 0.92 & 0.88 \\
LR meta, SVM simple & 0.71 & 0.76 & 0.71 & 0.85 & 0.92 & 0.88 \\
SVM meta, LR simple & 0.84 & 0.86 & 0.85 & 0.84 & 0.86 & 0.85 \\
\rowcolor{gray!15}SVM meta, SVM simple & 0.8 & 0.82 & 0.81 & 0.85 & 0.92 & 0.88 \\
\midrule
\multicolumn{7}{c}{\textbf{Data CERN}} \\
\midrule
\rowcolor{gray!15} Simple LR & 0.82 & 0.91 & 0.86 & 0.82 & 0.91 & 0.86\\
Simple SVM & 0.8 & 0.82 & 0.80 & 0.82 & 0.77 & 0.77 \\
LR meta, LR simple & 0.83 & 0.89 & 0.86 & 0.87 & 0.87 & 0.86 \\
LR meta, SVM simple & 0.83 & 0.89 & 0.86 & 0.85 & 0.85 & 0.85 \\
SVM meta, LR simple & 0.85 & 0.85 & 0.85 & 0.83 & 0.79 & 0.79 \\
SVM meta, SVM simple & 0.86 & 0.86 & 0.86 & 0.82 & 0.79 & 0.79 \\
\bottomrule
\end{tabularx}
\end{table*}
\begin{table*}[!htbp]
\centering
\caption{Model Performance Comparison Individual}
\label{tab:model_comparison_i}
\small
\setlength{\tabcolsep}{5pt} 
\begin{tabularx}{1\textwidth}{p{3.7cm} *{12}{>{\centering\arraybackslash}X}}
\toprule
Classifier & \multicolumn{3}{c}{What H1} & \multicolumn{3}{c}{What H2} & \multicolumn{3}{c}{Why H1} & \multicolumn{3}{c}{Why H2} \\
\cmidrule(lr){2-4} \cmidrule(lr){5-7} \cmidrule(lr){8-10} \cmidrule(lr){11-13}
& \makebox[2em]{Precision} & \makebox[2em]{Recall} & \makebox[2em]{F1} & \makebox[2em]{Precision} & \makebox[2em]{Recall} & \makebox[2em]{F1} & \makebox[2em]{Precision} & \makebox[2em]{Recall} & \makebox[2em]{F1} & \makebox[2em]{Precision} & \makebox[2em]{Recall} & \makebox[2em]{F1} \\
\addlinespace 
\midrule
\multicolumn{13}{c}{Data ACC} \\
\midrule
LR & 0.29 & 0.54 & 0.38 & 0.41 & 0.64 & 0.50 & 0.41 & 0.64 & 0.50 & 0.29 & 0.54 & 0.38 \\
SVM & 0.63 & 0.66 & 0.65 & 0.59 & 0.58 & 0.58 & 0.36 & 0.40 & 0.38 & 0.61 & 0.62 & 0.60 \\
LR meta, LR simple & 0.72 & 0.56 & 0.42 & 0.40 & 0.62 & 0.49 & 0.41 & 0.64 & 0.50 & 0.29 & 0.54 & 0.38 \\
LR meta, SVM simple & 0.50 & 0.54 & 0.44 & 0.39 & 0.56 & 0.46 & 0.41 & 0.64 & 0.50 & 0.29 & 0.54 & 0.38 \\
SVM meta, LR simple & 0.62 & 0.62 & 0.58 & 0.65 & 0.68 & 0.65 & 0.37 & 0.50 & 0.43 & 0.55 & 0.56 & 0.49 \\
\rowcolor{gray!15}SVM meta, SVM simple & 0.55 & 0.58 & 0.56 & 0.54 & 0.56 & 0.55 & 0.37 & 0.50 & 0.43 & 0.55 & 0.56 & 0.49 \\
\midrule
\multicolumn{1}{c}{Imbalance Mitigation}\\
\midrule
Self-Training & 0.77 & 0.60 & 0.49 & 0.79 & 0.69 & 0.60 & 0.41 & 0.62 & 0.49 & 0.76 & 0.55 & 0.41 \\
SMOTE & 0.78 & 0.64 & 0.57 & 0.71 & 0.69 & 0.63 & 0.53 & 0.62 & 0.53 & 0.60 & 0.55 & 0.44 \\
Non-Consensus & 0.79 & 0.60 & 0.51 & 0.81 & 0.67 & 0.61 & 0.41 & 0.60 & 0.49 & 0.76 & 0.55 & 0.45 \\
SMOTE + Non-Consensus & 0.79 & 0.64 & 0.58 & 0.70 & 0.67 & 0.62 & 0.41 & 0.60 & 0.49 & 0.76 & 0.55 & 0.45 \\
\midrule
\multicolumn{13}{c}{Data CERN} \\
\midrule
LR & 0.24 & 0.49 & 0.32 & 0.64 & 0.80 & 0.71 & 0.38 & 0.62 & 0.47 & 0.46 & 0.68 & 0.55 \\
SVM & 0.58 & 0.58 & 0.57 & 0.63 & 0.54 & 0.57 & 0.61 & 0.57 & 0.57 & 0.51 & 0.45 & 0.46 \\
LR meta, LR simple & 0.27 & 0.51 & 0.35 & 0.67 & 0.77 & 0.71 & 0.55 & 0.58 & 0.55 & 0.57 & 0.62 & 0.59 \\
LR meta, SVM simple & 0.27 & 0.51 & 0.35 & 0.67 & 0.77 & 0.71 & 0.51 & 0.54 & 0.52 & 0.60 & 0.63 & 0.61 \\
SVM meta, LR simple & 0.53 & 0.52 & 0.47 & 0.69 & 0.66 & 0.68 & 0.56 & 0.55 & 0.56 & 0.51 & 0.49 & 0.50 \\
\rowcolor{gray!15}SVM meta, SVM simple & 0.53 & 0.52 & 0.47 & 0.70 & 0.68 & 0.69 & 0.61 & 0.60 & 0.60 & 0.48 & 0.45 & 0.46\\
\midrule
\multicolumn{1}{c}{Imbalance Mitigation}\\
\midrule
Self-Training & 0.76 & 0.48 & 0.33 & 0.70 & 0.82 & 0.76 & 0.34 & 0.58 & 0.43 & 0.46 & 0.68 & 0.55 \\
SMOTE & 0.64 & 0.50 & 0.41 & 0.68 & 0.70 & 0.69 & 0.56 & 0.58 & 0.55 & 0.57 & 0.60 & 0.58 \\
Non-Consensus & 0.77 & 0.48 & 0.35 & 0.69 & 0.76 & 0.73 & 0.49 & 0.54 & 0.48 & 0.65 & 0.68 & 0.64 \\
SMOTE + Non-Consensus & 0.78 & 0.50 & 0.39 & 0.69 & 0.74 & 0.71 & 0.53 & 0.56 & 0.52 & 0.59 & 0.62 & 0.60 \\
\bottomrule
\end{tabularx}
\end{table*}
\begin{table}[!htbp]
\footnotesize
\centering
\caption{Class Distributions for ACC/CERN Datasets}
\label{tab:class_distributions}
\setlength{\tabcolsep}{4pt}
\begin{tabular}{l
>{\raggedright\arraybackslash}p{0.5cm}
>{\raggedright\arraybackslash}p{0.5cm}>{\raggedright\arraybackslash}p{0.5cm}>{\raggedright\arraybackslash}p{0.5cm}>{\raggedright\arraybackslash}p{0.5cm}>{\raggedright\arraybackslash}p{0.5cm}}
\toprule
\textbf{Task} & \multicolumn{3}{c}{\textbf{ACC (42 defeaters)}} & \multicolumn{3}{c}{\textbf{CERN (50 defeaters)}} \\
\cmidrule(lr){2-4} \cmidrule(lr){5-7}
& \textbf{H1} & \textbf{H2} & \textbf{Model} & \textbf{H1} & \textbf{H2} & \textbf{Model} \\
& \scriptsize (0,1,2) & \scriptsize (0,1,2) & \scriptsize (0,1,2) & \scriptsize (0,1,2) & \scriptsize (0,1,2) & \scriptsize (0,1,2) \\
\midrule
What & 0, 19, 23 & 0, 15, 27 & 0, 6, 36 & 0, 27, 23 & 0, 8, 42 & 2, 1, 47 \\
Why & 0, 15, 27 & 0, 20, 22 & 0, 2, 40 & 0, 21, 29 & 0, 16, 34 & 0, 4, 46 \\
Where & 10, 7, 25 & 1, 15, 26 & 19, 0, 23 & 2, 8, 40 & 3, 3, 44 & 13, 0, 37 \\
\bottomrule
\end{tabular}
\end{table}

\begin{table}[!htbp]
\footnotesize
\centering
\caption{Inter-Rater Agreement (Cohen’s Kappa) ACC/CERN}
\label{tab:kappa}
\setlength{\tabcolsep}{3pt}
\begin{tabular}{lcccc}
\toprule
\textbf{Dataset} & \textbf{Task} & \textbf{H1/H2} & \textbf{H1/Model} & \textbf{H2/Model} \\
\midrule
\multirow{3}{*}{ACC} & What & 0.02 & 0.34 & 0.34 \\
& Why & -0.50 & -0.09 & 0.10 \\
& Where & 0.33 & 0.33 & 0.27 \\
\midrule
\multirow{3}{*}{CERN} & What & -0.18 & 0.07 & -0.06 \\
& Why & -0.40 & -0.06 & -0.03 \\
& Where & 0.16 & 0.35 & 0.40 \\
\bottomrule
\end{tabular}
\end{table}

This section assesses classifier configurations for automated defeater evaluation, addressing RQ1 (accuracy and consistency with human raters) and RQ2 (subjectivity reduction) from Section \ref{subsec:Hypotheses}. To our knowledge, it is the first approach automating the evaluation of generated defeaters; for that, we tested Simple LR, SVM, and meta-classifiers (LR and SVM variants) on ``What'' and ``Why'' components against human annotations (H1, H2). The ``Where'' component was evaluated with the extraction logic explained in Subsection \ref{Subsec:Extraction Logic}. We show below the weighted F1-score results in Table \ref{tab:model_comparison} for general match against both raters and in Table \ref{tab:model_comparison_i} for individual matches, additionaly listed the imbalance mitigation strategies introduced in Subsection \ref{subsec:Model Selection and Training}; Table \ref{tab:class_distributions} and \ref{tab:kappa} show the agreement results and class distribution the SVM simple SVM meta model achieved.

\subsection{RQ1}
For RQ1, embedding-based classifier models demonstrate comparable consistency with human expert ratings, as evidenced by weighted F1 scores in Tables \ref{tab:model_comparison} and \ref{tab:model_comparison_i}. For the ACC dataset, the SVM meta, LR simple models achieve the highest General What F1 score of 0.85, with a strong What H2 Individual F1 of 0.65. Simple SVM excels with What H1 Individual F1 of 0.65 and Why H2 Individual F1 of 0.60. For the CERN dataset, Simple LR achieves a general What/Why F1 score of 0.86. In contrast, LR Meta and SVM Simple lead with a What H2 individual F1 score of 0.71 and a Why H2 individual F1 score of 0.61. SVM Meta and SVM Simple top the Why H1 individual F1 score at 0.60, which is balanced for both raters and shows individual accuracy. The SVM Meta and SVM Simple models achieve an F1 score of 0.84 across the ACC and CERN datasets, thereby also supporting RQ1, which demonstrates the accuracy of the artificial ratings.

\subsection{RQ2}
For RQ2, classifier models reduce the impact of subjectivity in defeater validation, particularly in cases of significant disagreement, as demonstrated by the F1 scores in Table \ref{tab:model_comparison_i}. Balanced ratings across reviewers highlight this effect. The imbalance mitigation method involving self-training with iteratively refined predictions does not equalise performance among raters. Synthetic Minority Oversampling Technique (SMOTE) with artificial minority labels performs better, but the remaining imbalance is worse than with the SVM Meta and SVM Simple approaches. Adding additional non-consensus defeaters does not improve the balance of performance against both reviewers.
The subjectivity problem of independent raters is evident, shown by the class distributions of dissensus data in Table \ref{tab:class_distributions}. A deeper analysis of the agreement rates for the best overall performance model (SVM Meta and SVM Simple) in Table \ref{tab:kappa} reveals further insights. The initial negative kappa values: ACC Why (-0.50), CERN Why (-0.40), and CERN What (-0.18) indicate worse-than-chance agreement in the human ratings. Using the automated review approach incorporating the model improved the values by approximately 40\%. The 'Where' component did not improve much but is included to provide a complete overview. It is not representative due to its static filter mechanism, which was explained in Subsection \ref{Subsec:Extraction Logic}, not considering general system descriptions.

\subsection{Result Discussion}
These findings suggest that embedding-based classifiers can mitigate the scalability and subjectivity issues of manual defeater validation, as outlined in Section \ref{sec:Introduction}. By leveraging structured domain-specific embeddings, the models capture nuances relevant to automotive (ACC) and nuclear (CERN) assurance cases (ACs) \cite{viger_ai-supported_2024}, enhancing defeater evaluation shown by the results in Table \ref{tab:model_comparison} and \ref{tab:model_comparison_i}. However, the models lack human-like contextual understanding, occasionally missing critical domain-specific information (e.g., unmodeled risks in complex ACs) due to limitations in embedding patterns. This is noticeable in lower Individual F1 scores (e.g., 0.43 for ACC Why H1 from Table \ref{tab:model_comparison_i}), where nuanced defeater evaluation requires deeper contextual insight. The size of the validation datasets with Dissensus Defeaters is limited. The performance of the models could be statistically supported by larger validation sets.

\section{Conclusion}\label{sec:Con}
Results from the model comparison Table \ref{tab:model_comparison} confirm that the models, trained on embeddings from the domains, can predict defeater quality and reflect expert tendencies, addressing Research Question 1. The meta-classifier integrates assurance case features, thereby enhancing contextual understanding; yet, it still exhibits training imbalances. Despite balanced class weights, LR models, including the meta-classifier, tend to favor majority classes due to imbalanced data distributions, aligning predictions more with Reviewer 2’s labels than Reviewer 1’s. Notably, SVM models showed slight bias, demonstrating consistent performance against both reviewers in the individual performance comparison Table \ref{tab:model_comparison_i}.

For Research Question 2, this automated review approach reduces subjective uncertainties, as demonstrated by the improvement in inter-agreement rates shown in Table \ref{tab:kappa}. It provides an objective perspective on safety case defeaters by using semantic feature embeddings to achieve contextually informed evaluation. While this method has its advantages, like all supervised methods, it relies on expert-curated data and is not excluded from data change effects.\\

\textbf{Future Work} Advanced LLMs or improved embedding techniques, such as fine-tuning on domain-specific datasets, could address contextual understanding limitations, opening directions for research to enhance classifier performance and calibrate appearing generalization problems for the generation task. Furthermore, future work should explore hybrid expert/embedding-based approaches in the safety engineering process to enhance validation reliability and scalability for dynamic systems at runtime.
\bibliographystyle{IEEEtran}
\bibliography{export-data}

@article{Odu2025AutomaticIO,
  title={Automatic instantiation of assurance cases from patterns using large language models},
  author={Oluwafemi Odu and Alvine Boaye Belle and Song Wang and S{\`e}gla Kpodjedo and Timothy C. Lethbridge and Hadi Hemmati},
  journal={J. Syst. Softw.},
  year={2025},
  volume={222},
  pages={112353},
  url={https://api.semanticscholar.org/CorpusID:275911948}
}

@inproceedings{Keraghel2024RecentAI,
  title={Recent Advances in Named Entity Recognition: A Comprehensive Survey and Comparative Study},
  author={Imed Keraghel and Stanislas Morbieu and Mohamed Nadif},
  year={2024},
  url={https://api.semanticscholar.org/CorpusID:267060999}
}

@article{scikit-learn,
  title={Scikit-learn: Machine Learning in {P}ython},
  author={Pedregosa, F. and Varoquaux, G. and Gramfort, A. and Michel, V.
          and Thirion, B. and Grisel, O. and Blondel, M. and Prettenhofer, P.
          and Weiss, R. and Dubourg, V. and Vanderplas, J. and Passos, A. and
          Cournapeau, D. and Brucher, M. and Perrot, M. and Duchesnay, E.},
  journal={Journal of Machine Learning Research},
  volume={12},
  pages={2825--2830},
  year={2011}
}

@misc{Dataset,
  title = {AI-EA Implementation and Data},
  howpublished = {\url{https://zenodo.org/records/13368055}},
  note = {Accessed: 2025-07-10}
}

@techreport{Bloomfield-etal:defeaters24,
	AUTHOR = {Robin Bloomfield and Kate Netkachova and John Rushby},
	TITLE = {Defeaters and Eliminative Argumentation in {Assurance 2.0}},
	INSTITUTION = {Computer Science Laboratory, SRI International},
	YEAR = 2024,
	ADDRESS = {Menlo Park, CA},
	MONTH = may,
	NUMBER = {SRI-CSL-2024-01},
	NOTE = {Additional arxiv {2405.15800}}
}

@ARTICLE{2015_Assadi,
  author={Asaadi, Erfan and Denney, Ewen and Menzies, Jonathan and Pai, Ganesh J. and Petroff, Dimo},
  journal={Computer}, 
  title={Dynamic Assurance Cases: A Pathway to Trusted Autonomy}, 
  year={2020},
  volume={53},
  number={12},
  pages={35-46},
  doi={10.1109/MC.2020.3022030}}

@inproceedings{kurzidem_safesens_2023,
  author = {I. Kurzidem and S. Burton and P. Schleiss},
  title = {SafeSens - Uncertainty quantification of complex perception systems},
  booktitle = {2023 IEEE 26th Int. Conf. Intell. Transp. Syst. (ITSC)},
  year = {2023},
  pages = {5805--5810},
  publisher = {IEEE},
  doi = {10.1109/ITSC57777.2023.10422256}
}

@misc{turobov_using_2024,
  author = {A. Turobov and D. Coyle and V. Harding},
  title = {Using ChatGPT for thematic analysis},
  year = {2024},
  publisher = {arXiv},
  note = {arXiv:2405.08828},
  doi = {10.48550/arXiv.2405.08828},
  url = {http://arxiv.org/abs/2405.08828}
}

@book{everitt_encyclopedia_2005,
  author = {B. S. Everitt and D. C. Howell},
  title = {Encyclopedia of statistics in behavioral science},
  year = {2005},
  publisher = {John Wiley \& Sons},
  address = {Chichester, UK},
  isbn = {978-0-470-86080-9},
  doi = {10.1002/0470013192}
}

@inproceedings{varadarajan_enabling_2024,
	location = {Cham},
	title = {Enabling Theory-Based Continuous Assurance: A Coherent Approach with Semantics and Automated Synthesis},
	isbn = {9783031687389},
	doi = {10.1007/978-3-031-68738-9_13},
	shorttitle = {Enabling Theory-Based Continuous Assurance},
	pages = {173--187},
	booktitle = {Computer Safety, Reliability, and Security. {SAFECOMP} 2024 Workshops},
	publisher = {Springer Nature Switzerland},
	author = {Varadarajan, Srivatsan and Bloomfield, Robin and Rushby, John and Gupta, Gopal and Murugesan, Anitha and Stroud, Robert and Netkachova, Kateryna and Wong, Isaac Hong and Arias, Joaquín},
	editor = {Ceccarelli, Andrea and Trapp, Mario and Bondavalli, Andrea and Schoitsch, Erwin and Gallina, Barbara and Bitsch, Friedemann},
	date = {2024},
	langid = {english},
}

@techreport{goodenough_eliminative_2015,
  author = {J. B. Goodenough and C. B. Weinstock and A. Z. Klein},
  title = {Eliminative Argumentation: A Basis for Arguing Confidence in System Properties},
  year = {2015},
  institution = {Carnegie Mellon University},
  address = {Pittsburgh, PA, USA},
  number = {CMU/SEI-2015-TR-005},
  type = {Technical Report},
  url = {https://insights.sei.cmu.edu/documents/1248/2015_005_001_434813.pdf}
}

@techreport{graydon_examining_2025,
  author = {M. S. Graydon and S. M. Lehman},
  title = {Examining Proposed Uses of LLMs to Produce or Assess Assurance Arguments},
  year = {2025},
  institution = {NASA Peer Committee},
  address = {Washington, DC, USA},
  number = {20250001849},
  type = {Technical Report},
  url = {https://ntrs.nasa.gov/citations/20250001849}
}

@inproceedings{rushby_interpretation_2015,
  author = {J. Rushby},
  title = {The Interpretation and Evaluation of Assurance Cases},
  year = {2015},
  booktitle = {Proc. Comput. Sci.},
  publisher = {Elsevier},
  address = {San Francisco, CA, USA},
  url = {https://www.semanticscholar.org/paper/The-Interpretation-and-Evaluation-of-Assurance-Rushby/5196fbfbc98306ad2273fc61d420af8ce1452fe1}
}

@article{Kelly2004,
author = {Kelly, Tim and Weaver, Rob},
year = {2004},
month = {01},
pages = {},
title = {The goal structuring notation–a safety argument notation},
journal = {Proc Dependable Syst Networks Workshop Assurance Cases}
}

@inproceedings{ratiu_towards_2024,
	location = {Cham},
	title = {Towards an Argument Pattern for the Use of Safety Performance Indicators},
	isbn = {9783031687389},
	doi = {10.1007/978-3-031-68738-9_12},
	pages = {160--172},
	booktitle = {Computer Safety, Reliability, and Security. {SAFECOMP} 2024 Workshops},
	publisher = {Springer Nature Switzerland},
	author = {Ratiu, Daniel and Rohlinger, Tihomir and Stolte, Torben and Wagner, Stefan},
	editor = {Ceccarelli, Andrea and Trapp, Mario and Bondavalli, Andrea and Schoitsch, Erwin and Gallina, Barbara and Bitsch, Friedemann},
	date = {2024},
	langid = {english},
}

@INPROCEEDINGS{tihomir_automated_2024,
  author={Tihomir, Rohlinger},
  booktitle = {35th {IEEE} International Symposium on Software Reliability Engineering, {ISSRE} 2024 - Workshops, Tsukuba, Japan, October 28-31, 2024},
  title={Automated Interpretation of Fleet Incidents to Enable System Level Runtime Assurance}, 
  year={2024},
  volume={},
  number={},
  pages={91-94},
  doi={10.1109/ISSREW63542.2024.00053}}

@misc{sivakumar_gpt-4_2023,
  author = {M. Sivakumar and A. B. Belle and J. Shan and K. K. Shahandashti},
  title = {GPT-4 and Safety Case Generation: An Exploratory Analysis},
  year = {2023},
  publisher = {arXiv},
  note = {arXiv:2312.05696},
  doi = {10.48550/arXiv.2312.05696},
  url = {http://arxiv.org/abs/2312.05696}
}

@article{sivakumar_prompting_2024,
  author = {M. Sivakumar and A. B. Belle and J. Shan and K. K. Shahandashti},
  title = {Prompting GPT-4 to Support Automatic Safety Case Generation},
  journal = {Expert Syst. Appl.},
  volume = {255},
  pages = {124653},
  year = {2024},
  publisher = {Elsevier},
  doi = {10.1016/j.eswa.2024.124653},
  url = {https://www.sciencedirect.com/science/article/pii/S0957417424015203}
}

@inproceedings{khakzad_shahandashti_assessing_2024,
	location = {Lisbon Portugal},
	title = {Assessing the Impact of {GPT}-4 Turbo in Generating Defeaters for Assurance Cases},
	isbn = {9798400706097},
	url = {https://dl.acm.org/doi/10.1145/3650105.3652291},
	doi = {10.1145/3650105.3652291},
	eventtitle = {{FORGE} '24: 2024 {IEEE}/{ACM} First International Conference on {AI} Foundation Models and Software Engineering},
	pages = {52--56},
	booktitle = {Proceedings of the 2024 {IEEE}/{ACM} First International Conference on {AI} Foundation Models and Software Engineering},
	publisher = {{ACM}},
	author = {Khakzad Shahandashti, Kimya and Sivakumar, Mithila and Mohajer, Mohammad Mahdi and Boaye Belle, Alvine and Wang, Song and Lethbridge, Timothy},
	urldate = {2025-04-09},
	date = {2024-04-14},
	langid = {english},
}

@article{Chawla2002SMOTE,
author = {Chawla, Nitesh and Bowyer, Kevin and Hall, Lawrence and Kegelmeyer, W.},
year = {2002},
month = {06},
pages = {321-357},
title = {SMOTE: Synthetic Minority Over-sampling Technique},
volume = {16},
journal = {J. Artif. Intell. Res. (JAIR)},
doi = {10.1613/jair.953}
}

@inproceedings{yarowsky-1995-unsupervised,
    title = "Unsupervised Word Sense Disambiguation Rivaling Supervised Methods",
    author = "Yarowsky, David",
    booktitle = "33rd Annual Meeting of the Association for Computational Linguistics",
    month = jun,
    year = "1995",
    address = "Cambridge, Massachusetts, USA",
    publisher = "Association for Computational Linguistics",
    url = "https://aclanthology.org/P95-1026/",
    doi = "10.3115/981658.981684",
    pages = "189--196"
}

@inproceedings{devlin_bert_2019,
	location = {Minneapolis, Minnesota},
	title = {{BERT}: Pre-training of Deep Bidirectional Transformers for Language Understanding},
	url = {https://aclanthology.org/N19-1423/},
	doi = {10.18653/v1/N19-1423},
	shorttitle = {{BERT}},
	eventtitle = {{NAACL}-{HLT} 2019},
	pages = {4171--4186},
	booktitle = {Proceedings of the 2019 Conference of the North American Chapter of the Association for Computational Linguistics: Human Language Technologies, Volume 1 (Long and Short Papers)},
	publisher = {Association for Computational Linguistics},
	author = {Devlin, Jacob and Chang, Ming-Wei and Lee, Kenton and Toutanova, Kristina},
	editor = {Burstein, Jill and Doran, Christy and Solorio, Thamar},
	urldate = {2025-01-26},
	date = {2019-06},
}

@inproceedings{viger_ai-supported_2024,
	title = {{AI}-Supported Eliminative Argumentation: Practical Experience Generating Defeaters to Increase Confidence in Assurance Cases},
	url = {https://ieeexplore.ieee.org/document/10771339/},
	doi = {10.1109/ISSRE62328.2024.00035},
	shorttitle = {{AI}-Supported Eliminative Argumentation},
	eventtitle = {2024 {IEEE} 35th International Symposium on Software Reliability Engineering ({ISSRE})},
	year = {2024},
    pages = {284--294},
	booktitle = {2024 {IEEE} 35th International Symposium on Software Reliability Engineering ({ISSRE})},
	author = {Viger, Torin and Murphy, Logan and Diemert, Simon and Menghi, Claudio and Joyce, Jeff and Di Sandro, Alessio and Chechik, Marsha},
	urldate = {2025-01-21},
	date = {2024-10},
	note = {{ISSN}: 2332-6549},
}

@inproceedings{millet_assurance_2023,
	location = {Cham},
	title = {Assurance Case Arguments in the Large: The {CERN} {LHC} Machine Protection System},
	isbn = {978-3-031-40923-3},
	pages = {3--10},
	booktitle = {Computer Safety, Reliability, and Security},
	publisher = {Springer Nature Switzerland},
	author = {Millet, Laure and Diemert, Simon and Rees, Chris and Viger, Torin and Chechik, Marsha and Menghi, Claudio and Joyce, Jeffrey},
	editor = {Guiochet, Jérémie and Tonetta, Stefano and Bitsch, Friedemann},
	date = {2023},
}

@inproceedings{gohar_codefeater_2024,
	location = {New York, {NY}, {USA}},
	title = {{CoDefeater}: Using {LLMs} To Find Defeaters in Assurance Cases},
	isbn = {9798400712487},
	url = {https://dl.acm.org/doi/10.1145/3691620.3695296},
	doi = {10.1145/3691620.3695296},
	series = {{ASE} '24},
	shorttitle = {{CoDefeater}},
	pages = {2262--2267},
	booktitle = {Proceedings of the 39th {IEEE}/{ACM} International Conference on Automated Software Engineering},
	publisher = {Association for Computing Machinery},
	author = {Gohar, Usman and Hunter, Michael C. and Lutz, Robyn R. and Cohen, Myra B.},
	urldate = {2025-01-10},
	date = {2024-10-27},
}

@inproceedings{first_baldur_2023,
	location = {New York, {NY}, {USA}},
	title = {Baldur: Whole-Proof Generation and Repair with Large Language Models},
	isbn = {9798400703270},
	url = {https://dl.acm.org/doi/10.1145/3611643.3616243},
	doi = {10.1145/3611643.3616243},
	series = {{ESEC}/{FSE} 2023},
	shorttitle = {Baldur},
	pages = {1229--1241},
	booktitle = {Proceedings of the 31st {ACM} Joint European Software Engineering Conference and Symposium on the Foundations of Software Engineering},
	publisher = {Association for Computing Machinery},
	author = {First, Emily and Rabe, Markus N. and Ringer, Talia and Brun, Yuriy},
	urldate = {2025-01-10},
	date = {2023-11-30},
}

@inproceedings{chen_use_2023,
	location = {Hannover, Germany},
	title = {On the Use of {GPT}-4 for Creating Goal Models: An Exploratory Study},
	rights = {https://doi.org/10.15223/policy-029},
	isbn = {9798350326918},
	url = {https://ieeexplore.ieee.org/document/10260905/},
	doi = {10.1109/REW57809.2023.00052},
	shorttitle = {On the Use of {GPT}-4 for Creating Goal Models},
	eventtitle = {2023 {IEEE} 31st International Requirements Engineering Conference Workshops ({REW})},
	pages = {262--271},
	booktitle = {2023 {IEEE} 31st International Requirements Engineering Conference Workshops ({REW})},
	publisher = {{IEEE}},
	author = {Chen, Boqi and Chen, Kua and Hassani, Shabnam and Yang, Yujing and Amyot, Daniel and Lessard, Lysanne and Mussbacher, Gunter and Sabetzadeh, Mehrdad and Varró, Dániel},
	urldate = {2025-01-05},
	date = {2023-09},
}

@article{DBLP:journals/corr/abs-1810-04805,
  author    = {Jacob Devlin and
               Ming{-}Wei Chang and
               Kenton Lee and
               Kristina Toutanova},
  title     = {{BERT:} Pre-training of Deep Bidirectional Transformers for Language
               Understanding},
  journal   = {CoRR},
  volume    = {abs/1810.04805},
  year      = {2018},
  url       = {http://arxiv.org/abs/1810.04805},
  archivePrefix = {arXiv},
  eprint    = {1810.04805},
  bibsource = {dblp computer science bibliography, https://dblp.org}
}

@techreport{ansi_ul_4600_2023,
  author = {{ANSI/UL}},
  title = {4600:2023 Standard for Safety for the Evaluation of Autonomous Products},
  year = {2023},
  publisher = {Underwriters Laboratories Standards \& Engagement},
  address = {Northbrook, IL, USA},
  note = {Standard, 3rd ed.},
  url = {https://www.shopulstandards.com/ProductDetail.aspx?productId=UL4600_3_S_20230317}
}

@techreport{iso_26262_1_2018,
  author = {{ISO}},
  title = {26262-1:2018 Road vehicles -- Functional safety -- Part 1: Vocabulary},
  year = {2018},
  publisher = {International Organization for Standardization},
  address = {Geneva, Switzerland},
  note = {Standard},
  isbn = {978-0-7381-9646-6},
  url = {https://www.iso.org/standard/68383.html}
}

@techreport{ISO21448,
  author = {{ISO}},
  title = {{21448:2022 Road vehicles -- Safety of the intended functionality (SOTIF)}},
  year = {2022},
  publisher = {International Organization for Standardization},
  address = {Geneva, Switzerland},
  note = {Standard},
  isbn = {978-0-7381-9646-6},
  url = {https://www.iso.org/standard/77490.html}
}

@inproceedings{vaswani_attention_2017,
	title = {Attention is All you Need},
	volume = {30},
	url = {https://papers.nips.cc/paper_files/paper/2017/hash/3f5ee243547dee91fbd053c1c4a845aa-Abstract.html},
	booktitle = {Advances in Neural Information Processing Systems},
	publisher = {Curran Associates, Inc.},
	author = {Vaswani, Ashish and Shazeer, Noam and Parmar, Niki and Uszkoreit, Jakob and Jones, Llion and Gomez, Aidan N and Kaiser, Lukasz and Polosukhin, Illia},
	urldate = {2024-12-16},
	date = {2017},
}

@book{koopman_how_2022,
  author = {P. Koopman},
  title = {How Safe Is Safe Enough? Measuring and Predicting Autonomous Vehicle Safety},
  year = {2022},
  publisher = {Carnegie Mellon University},
  address = {Pittsburgh, PA, USA},
  isbn = {979-8-8462-5124-3}
}

\end{document}